# Properties of an adjustable quarter-wave system under condition of multiple beam interference

Evelina A. Bibikova[1,2] and Nataliya D. Kundikova[1,2*]

[1] *Institute of Electrophysics of Urals Branch of RAS, Ekaterinburg, Russia*

[2] *South Urals State University, Chelyabinsk, Russia*

*Corresponding author: knd@susu.ac.ru

The polarimetric properties of an adjustable two plate quarter-wave system have been investigated. Multiple beam interference within single wave-plates has been taken into account. It has been shown that different adjustments are needed to produce left-handed and right-handed circular polarized coherent light. Laser light polarization conversion by the systems consisting of two birefringent mica plates has been investigated experimentally. The high-quality circularly polarized light with the intensity-related ellipticity higher than 0.99 has been produced. © 2012 Optical Society of America

*OCIS codes:* 260.5430, 230.2090, 230.3720, 260.1440

## 1. Introduction

Various experimental methods and applications of polarizing optics are frequently based on the use of circularly polarized light. Circularly polarized light is used for quality control of the circular-polarization-maintaining optical fiber [1], rotation of isotropic transparent microparticles in optical tweezers [2], precise Muller polarimetry [3] and for improvement of imaging through a turbulent medium [4]. An increase in interest to circularly polarized light is also connected with ability of sea animals to distinguish between left-circularly and right-circularly polarized light [5].

A quarter-wave plate is commonly used for circularly polarized light production. A quarter-wave plate is a linear retarder with the phase retardance between two orthogonal linear eigenpolarizations equal to $\pi/2$. This plate are usually made from birefringent materials such as crystal quartz or mica. In the case of incoherent light the phase retardance $\Gamma_0$ of a wave-plate [6] is given by:

$$\Gamma_0 = \frac{2\pi \Delta n d}{\lambda}, \qquad (1)$$

where $\Delta n$ is the birefringence of plate material and $d$ is the plate thickness. Really, it is difficult to make a quartz wave-plate with the phase retardance equal to 90°. It is easy to show that the inaccuracies of the wave-plate thickness equal to $\sim 10$ $\mu$m results in the phase retardance uncertainties approximately equal to 1° for the birefringence of quartz $\Delta n = 0.00903$ [6] at wavelength $\lambda = 0.65$ $\mu$m. A mica quarter-wave plate can be made by splitting a mica crystal into thin plates [6], but it is obvious, that the phase retardance of any thin plate will be equal to 90° only by chance.

The transformation of incoherent linearly polarized light by a system of two arbitrary oriented wave-plates has been considered in paper [7] on the basis of the Jones calculus. This two wave-plate polarization system is an elliptic retarder [8], its eigenpolarizations are ellipses, the major axes of the ellipses are perpendicular. According to [7] this system is equivalent to a system consisting of a linear retarder with the effective phase retardation following by an optical rotator. The adjustable polarization system consisted of two wave-plates with almost arbitrary phase shifts can acquire the effective phase retardance equal to $\pi/2$ by tuning the angle $\varphi$ between the slow axes of the wave-plates. This system can act as a quarter-wave plate and can be used for circularly polarized light production. It should be stressed, that it is possible to adjust this quarter-wave system for any light wavelength by adjusting the angle $\varphi$ between the slow axes [9].

The high coherence of light leads to multiple beam reflection interference and influences the plate retardance $\Gamma$. As a result the relative amplitude transmission coefficient $F$ of a wave-plate is not equal to unit [10]. That is why it is possible to consider the wave-plate as a combination of a linear retarder and a partial polarizer. It has been shown theoretically [10], that the dependence of the phase retardance on the wave-plates thickness $\Gamma(d)$ oscillates under the multiple beam interference. The dependence $\Gamma(d)$ shows that the zero order quartz quarter-wave plate is realized at twenty various plate thicknesses in the range of 3 $\mu$m. According to [10] if the accuracy of the plate thickness is equal to 0.1 $\mu$m, than the uncertainty of the phase retardance will be equal to 10°. These estimations show that it is practically impossible to make the high quality quartz quarter-wave plate for the coherent light polarization transformation. The effects of interference within a simple wave-plate have been investigated in papers [10,11]. The properties of a compound quarter-wave plate has been investigated theoretically for coherent light in paper [12]. The multiple beam interference within single plates made of different materials has been taken into account in paper [12].

In this paper, the polarimetric properties of an adjustable two plates quarter-wave system with the plates made of the same material are investigated theoretically and experimentally under conditions of multiple beam reflections interference.

## 2. Coherent light propagation through an adjustable quarter-wave system consisting of two plates

Let us use the Jones calculus [3] to describe coherent light propagation through an adjustable quarter-wave system consisting of two plates. As it was already stressed, a single wave-plate under the multiple beam interference can be described as a combination of a linear retarder and a partial polarizer, therefore a Jones matrix of such wave-plate is represented as follows:

$$\mathbf{T}(\Gamma, F) = \mathbf{T}^{\mathrm{LP}}(\Gamma) \times \mathbf{T}^{\mathrm{LA}}(F). \tag{2}$$

Here $\mathbf{T}^{\mathrm{LP}}(\Gamma)$ is a Jones matrix of a linear retarder with the phase retardance $\Gamma$:

$$\mathbf{T}^{\mathrm{LP}}(\Gamma) = \begin{pmatrix} \exp(-i\Gamma/2) & 0 \\ 0 & \exp(i\Gamma/2) \end{pmatrix} \tag{3}$$

and $\mathbf{T}^{\mathrm{LA}}(F)$ is a Jones matrix of a partial polarizer with the coefficient of linear amplitude dichroism $F$:

$$\mathbf{T}^{\mathrm{LA}}(F) = \begin{pmatrix} 1 & 0 \\ 0 & F \end{pmatrix}. \tag{4}$$

Thus $\mathbf{T}(\Gamma, F)$ can be written in the following form:

$$\mathbf{T}(\Gamma, F) = \begin{pmatrix} \exp(-i\Gamma/2) & 0 \\ 0 & F\exp(i\Gamma/2) \end{pmatrix}. \tag{5}$$

Let us show that a two plates adjustable system can transform linearly polarized light into circularly polarized one under conditions of the multiple beam interference, namely it can act as a quarter-wave plate. Let the $x$- and $y$-axes of a reference system are parallel to the slow and the fast axes of the first wave-plate. The slow axis of the second plate makes an angle $\varphi$ with the $x$-axis. The light wave propagates in a positive direction of the $z$-axis towards to an observer. A Jones matrix of the two plates adjustable system can be expressed in the following form:

$$\mathbf{W}(\varphi, \Gamma_1, \Gamma_2, F_1, F_2) = \mathbf{R}(-\varphi)\mathbf{T}(\Gamma_2, F_2)\mathbf{R}(\varphi)\mathbf{T}(\Gamma_1, F_1). \tag{6}$$

Here $\Gamma_1$ and $\Gamma_2$ are the phase retardances, $F_1$ and $F_2$ are the coefficient of linear amplitude dichroism of the first and second wave-plates, $\mathbf{R}(\pm\varphi)$ is a rotation matrix. In the sequel the parameters of single wave-plates are assumed to be identical: $\Gamma_1 = \Gamma_2 = \Gamma$ and $F_1 = F_2 = F$.

Let us determine the values of the adjusting angles $\varphi$, required for the transformation of linearly polarized light into circularly polarized light. Let $\alpha$ is an azimuth of incident linear polarized light with respect to the $x$-axis. The matrix equation describing this transformation looks like:

$$\rho \exp(i\psi) \cdot \begin{pmatrix} 1 \\ \sigma i \end{pmatrix} = \mathbf{W}(\varphi, \Gamma, F) \cdot \begin{pmatrix} \cos\alpha \\ \sin\alpha \end{pmatrix}. \tag{7}$$

Here $\sigma = +1$ stands for right circularly polarized light, and $\sigma = -1$ stands for left circularly polarized light. Value $\psi$ is a common phase shift, $\rho$ characterizes the change of the whole light intensity. A set of four real equations with four unknown quantities $\alpha, \varphi, \rho$ and $\psi$ is derived from Eq. (7):

$$\begin{aligned}
\cos\varphi \cos\alpha \cos\Gamma + F \sin\varphi \sin\alpha &= \rho \cos\psi, \\
-\cos\varphi \cos\alpha \sin\Gamma &= \rho \sin\psi, \\
-F \sin\varphi \cos\alpha + F^2 \cos\varphi \sin\alpha \cos\Gamma &= -\sigma\rho \sin\psi, \\
F^2 \cos\varphi \sin\alpha \sin\Gamma &= \sigma\rho \cos\psi.
\end{aligned} \tag{8}$$

If $\sigma = -1$, the analytical solutions for the adjusting angle $\varphi$ and the azimuth $\alpha$ of incident linear polarized light are represented as

$$\mathrm{tg}\varphi^{L1} = \frac{-(F^2-1)\sin\Gamma + \sqrt{(F^2-1)^2 \sin^2\Gamma - 4F^2 \cos(2\Gamma)}}{2F}, \tag{9}$$

$$\mathrm{tg}\varphi^{L2} = \frac{-(F^2-1)\sin\Gamma - \sqrt{(F^2-1)^2 \sin^2\Gamma - 4F^2 \cos(2\Gamma)}}{2F}, \tag{10}$$

$$\mathrm{tg}\alpha^{L1} = \frac{-2\cos\Gamma}{(F^2+1)\sin\Gamma + \sqrt{(F^2-1)^2 \sin^2\Gamma - 4F^2 \cos(2\Gamma)}}. \tag{11}$$

$$\mathrm{tg}\alpha^{L2} = \frac{-2\cos\Gamma}{(F^2+1)\sin\Gamma - \sqrt{(F^2-1)^2 \sin^2\Gamma - 4F^2 \cos(2\Gamma)}}. \tag{12}$$

If $\sigma = +1$, these solutions look like:

$$\mathrm{tg}\varphi^{R1} = \frac{(F^2-1)\sin\Gamma + \sqrt{(F^2-1)^2 \sin^2\Gamma - 4F^2 \cos(2\Gamma)}}{2F}, \tag{13}$$

$$\mathrm{tg}\varphi^{R2} = \frac{(F^2 - 1)\sin\Gamma - \sqrt{(F^2 - 1)^2 \sin^2\Gamma - 4F^2 \cos(2\Gamma)}}{2F}, \tag{14}$$

$$\mathrm{tg}\alpha^{R1} = \frac{2\cos\Gamma}{(F^2 + 1)\sin\Gamma - \sqrt{(F^2 - 1)^2 \sin^2\Gamma - 4F^2 \cos(2\Gamma)}}. \tag{15}$$

$$\mathrm{tg}\alpha^{R2} = \frac{2\cos\Gamma}{(F^2 + 1)\sin\Gamma + \sqrt{(F^2 - 1)^2 \sin^2\Gamma - 4F^2 \cos(2\Gamma)}}. \tag{16}$$

It follows from Eq. (9) - (16), that it is possible to produce both left and right circularly polarized light at two various values of the adjusting angle $\varphi$, setting the corresponding azimuth of linear polarized incident light $\alpha$. It should be stressed, that the influence of the multiple beam interference leads to the impossibility to use the same configuration of the two plate adjustable system to produce both right and left circularly polarized light. It differs from the case of incoherent light [7].

It follows from Eq.(9) – (16), that $\varphi^{L1} = -\varphi^{R2}$ and $\varphi^{L2} = -\varphi^{R1}$. Each quarter-wave system for the linear polarized light transformation into circularly polarized light requires the azimuth $\alpha$ of incident linear polarized light to be determined by Eq. (11), Eq. (12), Eq. (15) and Eq. (16). The comparison of the Eq. (11), Eq. (12), Eq. (15) and Eq. (16) gives a following simple relation between these azimuths: $\alpha^{L1} = -\alpha^{R2}$ and $\alpha^{L2} = -\alpha^{R1}$.

It can be seen from Fig. 1, that $\varphi^{L1} = \varphi^{R1}$ and $\varphi^{L2} = -\varphi^{R2}$ only if $F = 1$. If the coifficient $F \neq 1$ the angle $\varphi$ should be changed in order to obtain right circularly polarized light. If the value of $F$ varies from 0.9 to 1.1, than the value of $\Delta\varphi = |\varphi^{L1} - \varphi^{R2}|$ will vary in the range up to 7° to obtain right circularly polarized light. It can be seen from Fig. 2, that $\alpha^{L1} - \alpha^{R1} = 90°$ and $\alpha^{L2} - \alpha^{R2} = 90°$, only if $F = 1$. If the value of $F \neq 1$ than $\alpha^{L1} - \alpha^{R1} \neq 90°$ and $\alpha^{L2} - \alpha^{R2} \neq 90°$ are the functions of $F$.

It should be stressed that the adjustable quarter-wave system designed for left circular polarized light can not produced right circular polarized light, but only right-handed elliptically polarized light. The obtained right-handed ellipticity depends on the value of $F$ and the value of $\Gamma$. The Fig. 3 shows the calculated dependences of the maximal possible intensity-related right-handed ellipticity $e^I$ at the output of the adjustable quarter-wave system designed for left circular polarized light. Figure 3 clearly shows that in order to obtain high quality circular polarized light with ($e^I = 1$) the system should be adjust.

## 3. Experimental results

Mica wave plates were used for construction of adjustable quarter-wave systems. The phase retardance $\Gamma$ and the relative amplitude transmission coefficient $F$ of the wave-plates were measured with the accuracy of $\Delta\Gamma = 0.05°$ and $\Delta F = 0.001$ by the method described in paper [13]. The wave-plates with the phase retardance $50 \div 60°$ at the wavelength $\lambda = 632.8$ nm have been chosen for experimental investigation. The pairs of highly identical wave-plates have been selected for the experiment. The identity of the pair of the wave-plates was checked by the method similar to one described in paper [14]. The deviation in the phase retardance between the selected wave-plates in each pairs did not exceed $0.1°$.

The angles $\varphi^{L1}$ and $\varphi^{R1}$ were calculated for each pair of the identical wave-plates with the parameters $\Gamma$, $F$. For simplicity let us omit indexes 1 and 2, so that $\varphi^{L1} = \varphi^L$ and $\varphi^{R1} = \varphi^R$, and corresponding azimuths of the input polarized light $\alpha^{L1} = \alpha^L$ and $\alpha^{R1} = \alpha^R$. The parameters of wave-plates for each system and calculated values of parameters of adjustable systems (the adjusting angles $\varphi^L$, $\varphi^R$ and the azimuths of input polarized light $\alpha^L$, $\alpha^R$) are listed in Table 1. The wave-plates selected for the experiment were characterized by the various values of $F$, both close to the unit, and significantly different from the unit. It has allowed to observe the influence of the multiple beam interference on the polarimetric properties of the systems. The uncertainties of the parameters of the single wave-plates $\Delta\Gamma = 0.05°$ and $F = 0.001$ has led to the accuracy of calculating angels $\Delta\varphi^L$, $\Delta\varphi^R$, $\Delta\alpha^L$, $\Delta\alpha^R$ approximately equal to $0.1°$.

The quality of produced circular polarized light has been investigated. The experimental setup is shown in Fig. 4. The laser beam has been divided by a semitransparent mirror into two beams. The first beam has been used for the beam intensity control. The second beam propagated through a polarizer (Glan prism), the system under investigation and an analyzer. Variation of the analyzer azimuth angle has allowed us to measure beam intensity $I$ at the different analyzer azimuth angle $\beta^A$. The dependence $I(\beta^A)$ has shown that the two plate adjustable quarter-wave system can produce polarized light with the intensity-related ellipticity $e^I \geq 0.99$.

## 4. Conclusion

It has been shown theoretically and experimentally that an adjustable two plate quarter-wave system can be used for production of high quality circularly polarized coherent light. The intensity-related ellipticity can be higher than 0.99. Different ajdustments of an adjustable two plate quarter-wave system are required for production of left-handed and right-handed circularly polarized light.

## 5. Acknowledgments

This work has been performed within the framework of programs of fundamental research of the Physics Division of RAS "Fundamental problems of photonics and physics of new optical materials" (12-T-2-1003).

Table 1. The adjusting angles of two plate quarter-wave systems (the values of all angles are given in degrees).

| N | Parameters of wave-plates | | Adjusting angles | | Azimuths of input polarisation | |
|---|---|---|---|---|---|---|
|   | $F$ |   | $\varphi^L$ | $\varphi^R$ | $\alpha^L$ | $\alpha^R$ |
| 1 | 1.010 | 63.90 | 37.7 | 38.4 | -14.5 | 75.2 |
| 2 | 1.009 | 50.95 | 24.1 | 24.7 | -26.9 | 62.7 |
| 3 | 1.051 | 50.40 | 21.6 | 25.3 | -26.7 | 61.0 |
| 4 | 1.090 | 59.05 | 31.6 | 37.4 | -16.9 | 70.1 |
| 5 | 1.075 | 52.25 | 24.1 | 29.3 | -23.7 | 63.1 |

**List of Figure Captions**

Fig. 1. The adjusting angle $\varphi$ as a function of the coefficient $F$ ($\Gamma = 50°$).

Fig. 2. The azimuth of the incident linearly polarized light $\alpha$ as a function of the coefficient $F$ ($\Gamma = 50°$).

Fig. 3. An intensity-related ellipticity $e^I$ of right-handed elliptically polarized light after propagation through the system for production of left circularly polarized light as function of the value $F$ for the different value of $\Gamma$.

Fig. 4. The experimental setup for the determination of the circularly polarized light quality.

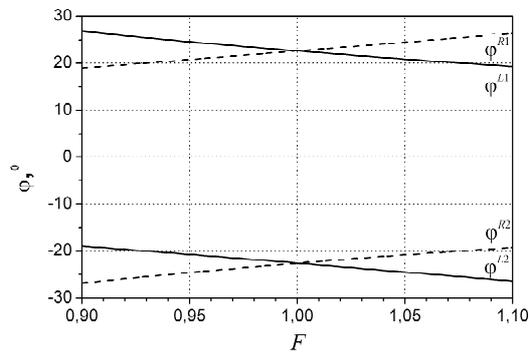

Fig. 1. The adjusting angle $\varphi$ as a function of the coefficient $F$ ($\Gamma = 50°$).

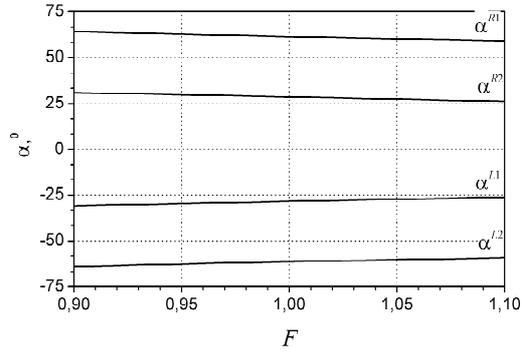

Fig. 2. The azimuth of the incident linearly polarized light $\alpha$ as a function of the coefficient $F$ ($\Gamma = 50°$).

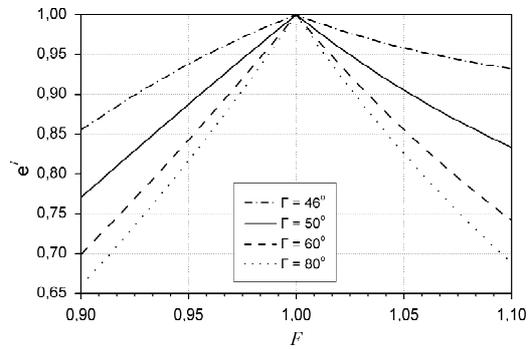

Fig. 3. An intensity-related ellipticity $e^I$ of right-handed ellipticlly polarized light after propagation through the system for production left circularly polarized light as function of the value $F$ for the different value of $\Gamma$.

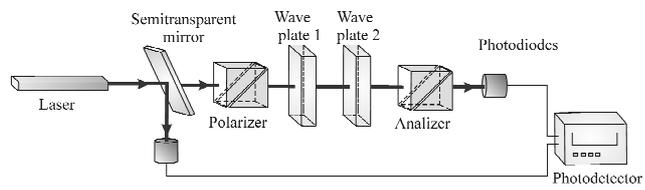

Fig. 4. The experimental setup for the determination of the circularly polarized light quality.